\documentclass[preprint,5p,number,twocolumn]{elsarticle}
\usepackage{lineno}

 \usepackage{graphics}
 \usepackage{graphicx}

\usepackage{amssymb}
\usepackage{amsthm}
\usepackage[FIGBOTCAP]{subfigure}
\usepackage{lineno}
\usepackage{amsmath,amssymb} 
\usepackage[pdftex]{hyperref}

\journal{Elsevier}

\begin{document}

\begin{frontmatter}



 \cortext[cor1]{Corresponding author}

\title{\Large\bf Test of \textit{Topmetal-${II}^-$} In Liquid Nitrogen For Cryogenic Temperature TPCs}

	 \author{Shuguang Zou}
	 \author[]{Yan Fan}
	\author[]{Mangmang An}
	\author[]{Chufeng Chen}
	\author[]{Guangming Huang}
	\author[]{Jun Liu}
	\author[]{ Hua Pei}
 \author{Xiangming Sun$^{*}$}
 \ead{xmsun@phy.ccnu.edu.cn}
 	\author[]{Ping Yang}
 	\author[]{Dong Wang}
 	\author[]{Le Xiao}
	\author[]{Zhen Wang}
	\author[]{Kai Wang}
	\author[]{ Wei Zhou}
	\address[ccnu]{PLAC, Key Laboratory of Quark $\&$ Lepton Physics (MOE), Central China Normal University, Wuhan, Hubei 430079, P.R.China}
\begin{abstract}
\textit{Topmetal-${II}^-$} is a highly pixelated direct charge sensor that contains a 72${\times}$72 pixel array of 83${\mu}$m pitch size. The key feature of \textit{Topmetal-${II}^-$} is that it can directly collect charges via metal nodes of each pixel to form two-dimensional images of charge cloud distributions. \textit{Topmetal-${II}^-$} was proved to measure charged particles without amplification at room temperature. To measure its performance at cryogenic temperature, a \textit{Topmetal-${II}^-$} sensor is embedded into a liquid nitrogen dewar. The results presented in this paper show that \textit{Topmetal-${II}^-$} can also operate well at this low temperature with a noise (ENC) of 12 e$^-$ lower than that at room temperature (13 e$^-$). From the noise perspective, \textit{Topmetal-${II}^-$} is a promising candidate for the next generation readout of liquid argon and xenon Time Projection Chamber (TPC) used in experiments searching for neutrinoless double beta decay and dark matter.

\end{abstract}

\begin{keyword}
Topmetal, Pixel, low noise sensor, cryogenic temperature
\end{keyword}

\end{frontmatter}

\section{Introduction}\label{Introduction}

	Over the years people have been interested in rare event experiments like neutrinoless double beta decay\cite{neutrinoless} and dark matter\cite{darkmatter} searches. Liquid argon and xenon are considered to be good media for low rate TPC detectors such as ICARUS\cite{ICARUS} and LANNDD\cite{liquid_argon_detector_LANNDD}. The traditional readouts of liquid argon or xenon TPCs are basically multi-wire electrodes. It is challenging to reduce the distance between two wires to a few hundreds microns or tens of microns for a large scale liquid argon or xenon detector\cite{largeScaleLiquidArgon}. 	

	On the other hand, a direct charge CMOS sensor with large pixel array is a good choice of readout for such low rate and large scale detectors, since CMOS sensors offer small resolution and high granularity. Timepix\cite{Timepix} chip is a good example of direct charge readout sensor that has a good performance at -125 $^\circ$C(148 K) with a noise of 99 e$^-$\cite{CoolTimepix}. Coupled with an aluminum mesh \cite{GridPix} where gas amplification occurs, Timepix can be applied in a dual phase argon TPC with high detection efficiency\cite{GridPixDualPhaseTPC}.

	We have designed a CMOS sensor named as \textit{Topmetal-${II}^-$} with rather low noise and high spatial resolution. It can be applied into a TPC detector as a charge collector to measure single electrons generated by alpha particles\cite{topmetal} at room temperature without any charge multiplier being necessary. This result prompted us to test if \textit{Topmetal-${II}^-$} can work at the temperature of liquid argon(83.8 K).	

	In this paper, we mainly study how \textit{Topmetal-${II}^-$} works in liquid nitrogen(77 K) and compare the performance of the sensor at room temperature and in liquid nitrogen. 
	
\section{\textit{Topmetal-${II}^-$}}
\begin{figure} [htbp]
		\centering
		\subfigure[]{\label{fig:subfig:a}
			\includegraphics[width=0.4\columnwidth]{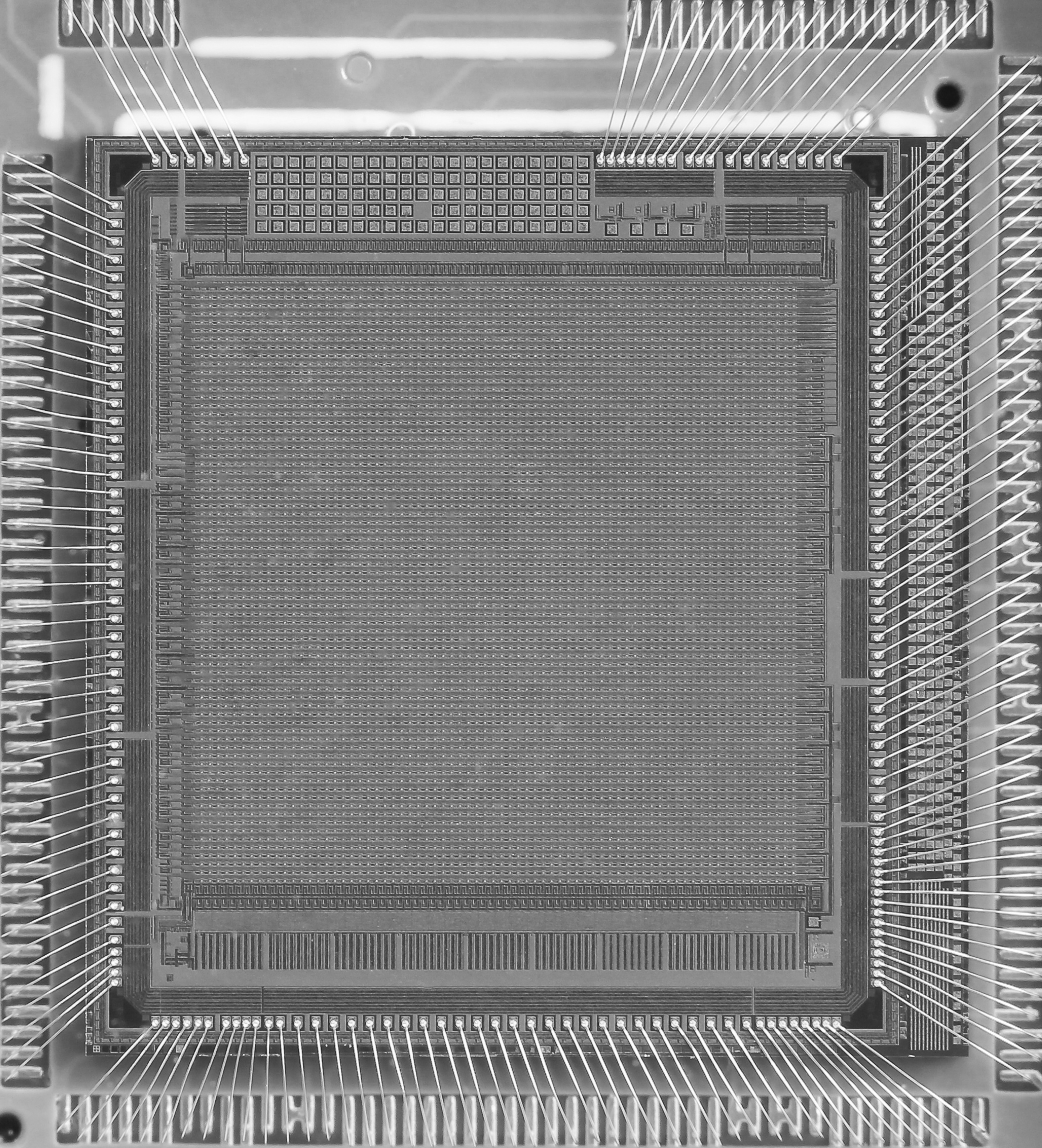}}
		\subfigure[]{\label{fig:subfig:b} 
			\includegraphics[width=0.4\columnwidth]{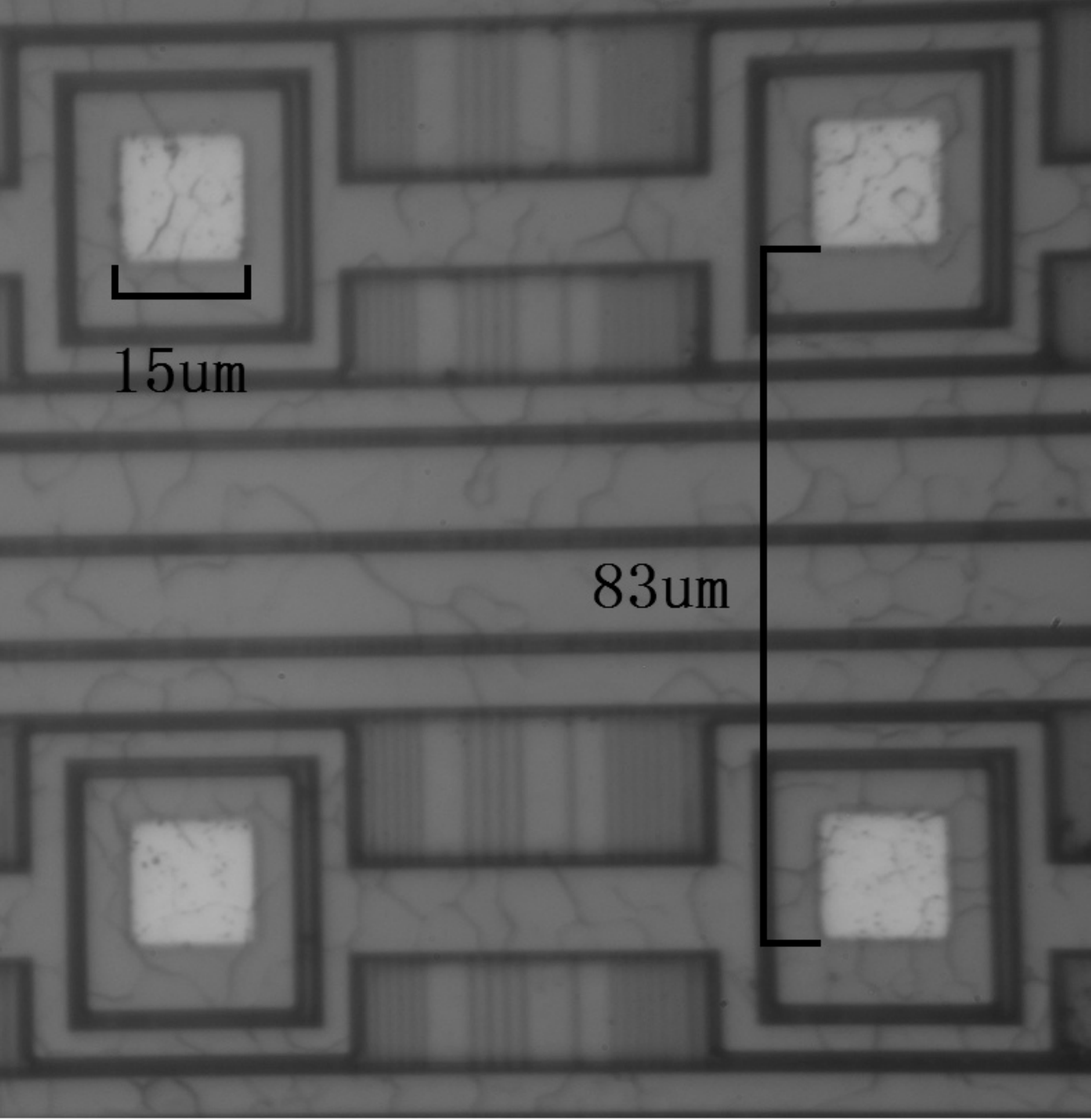}}
		\caption{(a) Photograph of a single \textit{Topmetal\textrm{-}$II^{-}$} sensor. It consists of 72$\times$72 pixels. (b) Zoom-in picture shows each pixel size is 83$\times$83 ${\mu}$m$^2$. The size of each charge collection electrode (in light color) is 15$\times$15 ${\mu}$m$^2$.}
		\label{Photograph of topmetal-II-}
		
	\end{figure}

	\textit{Topmetal-${II}^-$} is a direct charge sensor with high spatial resolution. Charges can be collected via metal nodes on each pixel when an electric field is applied above the top of sensor. A wire-bonded \textit{Topmetal-${II}^-$} sensor is shown in Figure. \ref{Photograph of topmetal-II-}. It consists of a 72${\times}$72 square pixel array. Each pixel size is 83$\times$83 ${\mu}$m$^2$.

\begin{figure} [htbp]
		\centering
		\includegraphics[width=1.0\columnwidth]{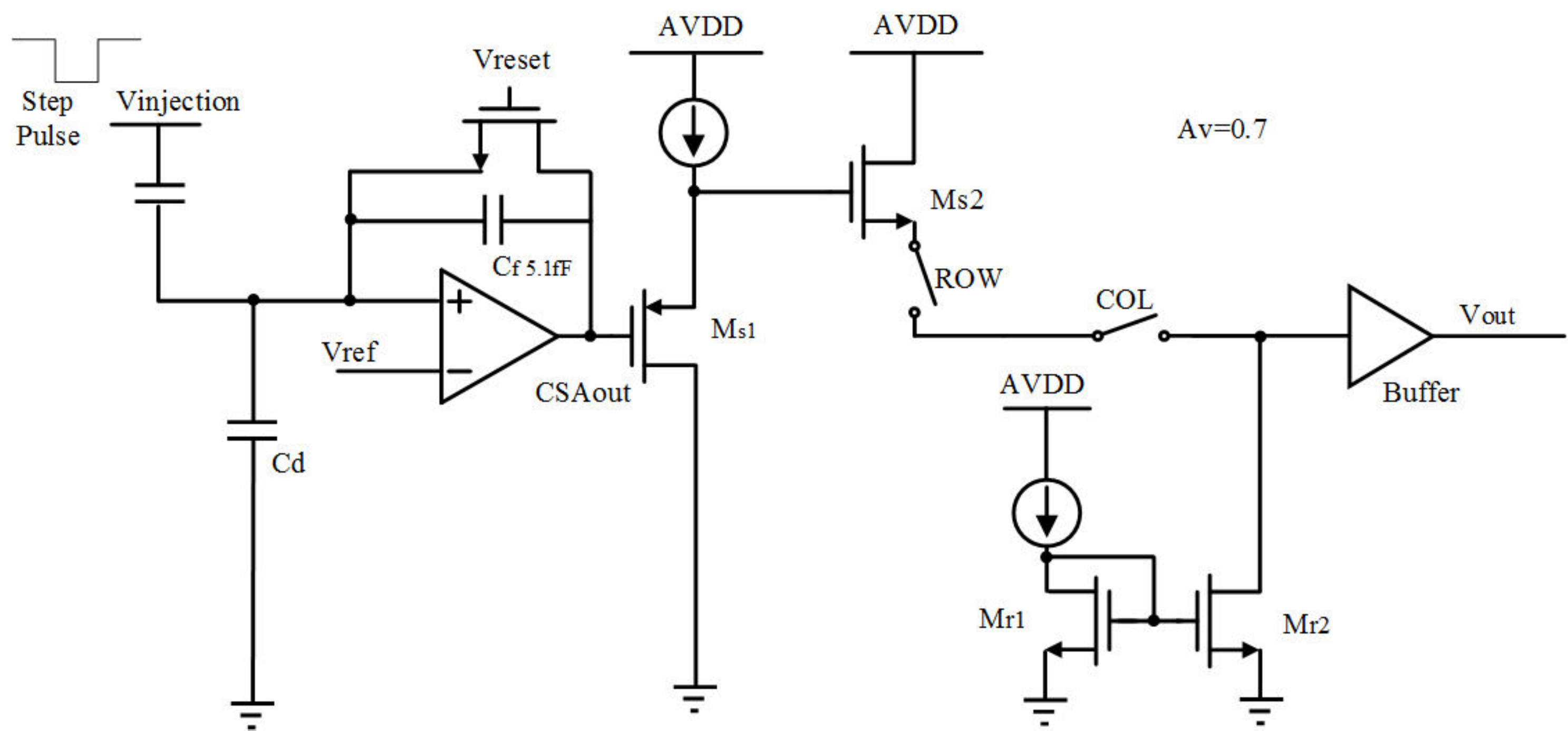}
		\centering
		\caption{Internal structure of analogue readout of a single pixel.}	
		\label{principle figure of analogue readout channel of Topmetal-II-}		
	\end{figure}

	The total structure of \textit{Topmetal-${II}^-$} has been described in an earlier paper\cite{topmetal}. Here the internal structure of single-pixel analogue readout is briefly presented in Figure. \ref{principle figure of analogue readout channel of Topmetal-II-}. Each pixel consists of a charge collection electrode, a charge sensitive amplifier (CSA), an analogue readout channel and a digital readout channel(not shown in figure). Analogue and digital readout channels operate independently. There is a guard ring at the periphery of the metal nodes of each pixel and the sensor's performance is measured by injecting pulse into the CSA through the guard ring. The capacitance between the guard ring and top metal of pixels ($C_d$) extracted by IC design software is about 5.5 fF.

\section{Test Results of \textit{Topmetal-${II}^-$}}
	In order to compare the performance of \textit{Topmetal-${II}^-$} at room temperature and cryogenic temperature, the operating setup of this sensor embedded inside liquid nitrogen has to be well designed. \textit{Topmetal-${II}^-$} sensor is connected to a PCB board, the PCB board is fixed on a lifting platform. We have no special protection on \textit{Topmetal-${II}^-$} sensor. The sensor is slowly moved into a liquid nitrogen dewar under the surface level about 4cm. After 15 minutes, the sensor is powered on and configured, then the signal is observed through an oscilloscope. This process is repeated many times. Three sensors are observed before and after the experiment using microscope, no significant difference is found.
	
	We first tested the performance of \textit{Topmetal-${II}^-$} at room temperature in ambient air and then put it into a liquid nitrogen dewar to measure its performance. The following results show that \textit{Topmetal-${II}^-$} works properly in liquid nitrogen and performs better than at room temperature with lower electronic noise for a single pixel in the absence of drift field.

	\subsection{Decay time constant}
		
		For measuring decay time constant, a square wave of 200 mV($\pm$100 mV) peak-to-peak amplitude is applied to the guard ring (internal test circuits of the sensor) of \textit{Topmetal-${II}^-$} in the absence of drift field. An equivalent charge of $C_{d}\times$200 mV=6.8$\times$10$^3$ $e^{-}$ is injected to each pixel, where positive equivalent charges are injected at rising edges of the square wave and negative equivalent charges are injected at falling edges. The sensor chip can be tested at a single pixel level and the result is shown in Figure. \ref{data of single pixel signal and after trapezoidal filter}. 
		
		\begin{figure} [htbp]
			\centering
			\includegraphics[width=1.0\columnwidth]{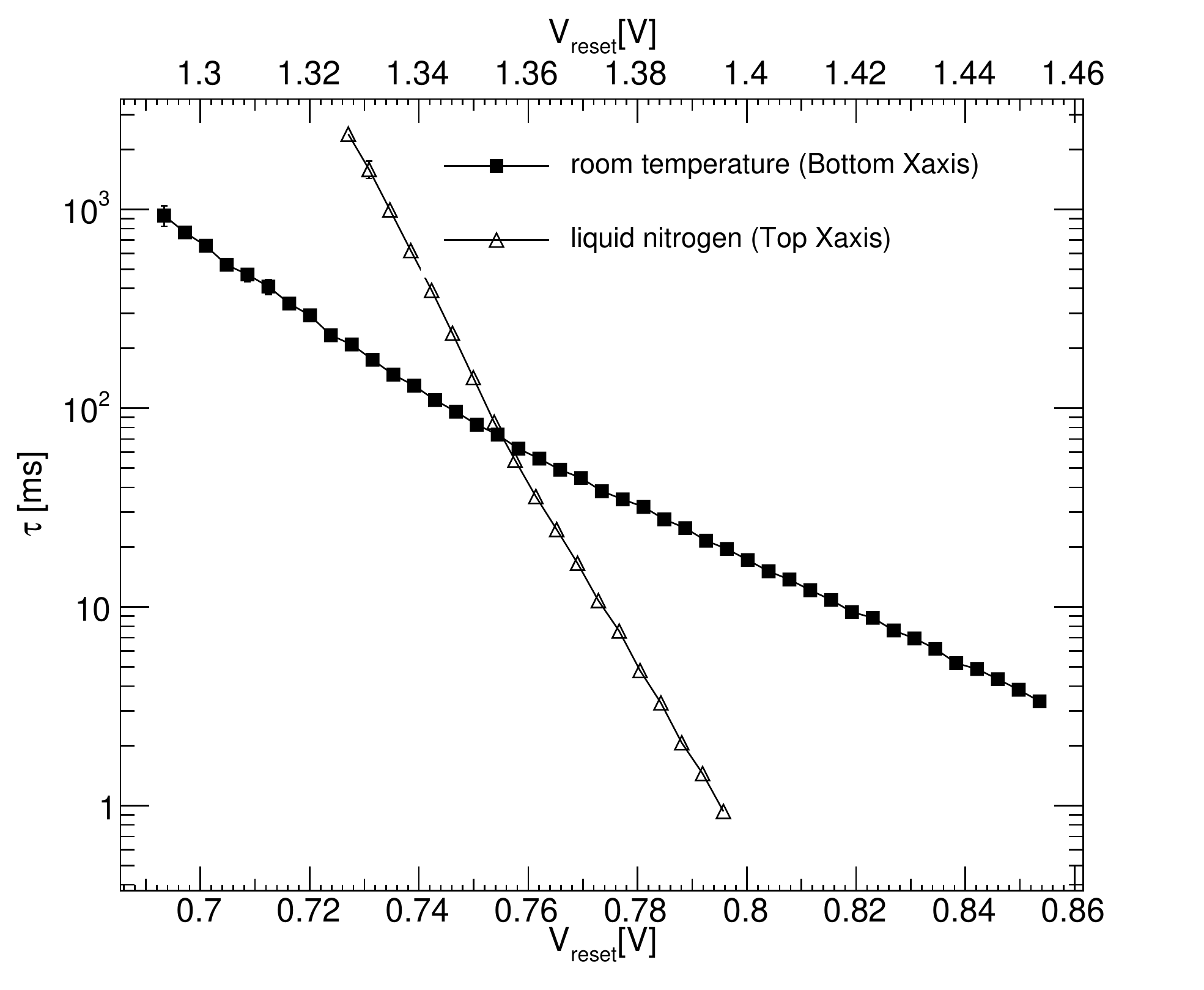}
			\centering
			\caption{A single pixel's reset voltage ($V_{reset}$) dependence of decay time constant of \textit{Topmetal-${II}^-$} at room temperature (solid square) and in liquid nitrogen (triangle). The sensor is proved to be able to work in liquid nitrogen with higher reset voltage (${V_{reset}}$) than that at room temperature.}	
			\label{Decay constant vs Vreset of single pixel at room temperature and in liquid nitrogen}		
		\end{figure}	
		
		The decay time constant of a single \textit{Topmetal-${II}^-$} pixel varies with temperature, thus different reset voltages $V_{reset}$ should be applied at room temperature and in liquid nitrogen to measure the variations of decay time constants with $V_{reset}$ and be adjusted to achieve similar desired mean values. In liquid nitrogen, \textit{Topmetal-${II}^-$} sensor can work with higher reset voltage (${V_{reset}}$) than that at room temperature. As shown in Figure. \ref{Decay constant vs Vreset of single pixel at room temperature and in liquid nitrogen}, the decay time constant of a pixel changes much faster with $V_{reset}$ in liquid nitrogen than at room temperature. 
	
		\begin{figure} [htbp]
			\centering
			\includegraphics[width=1.0\columnwidth]{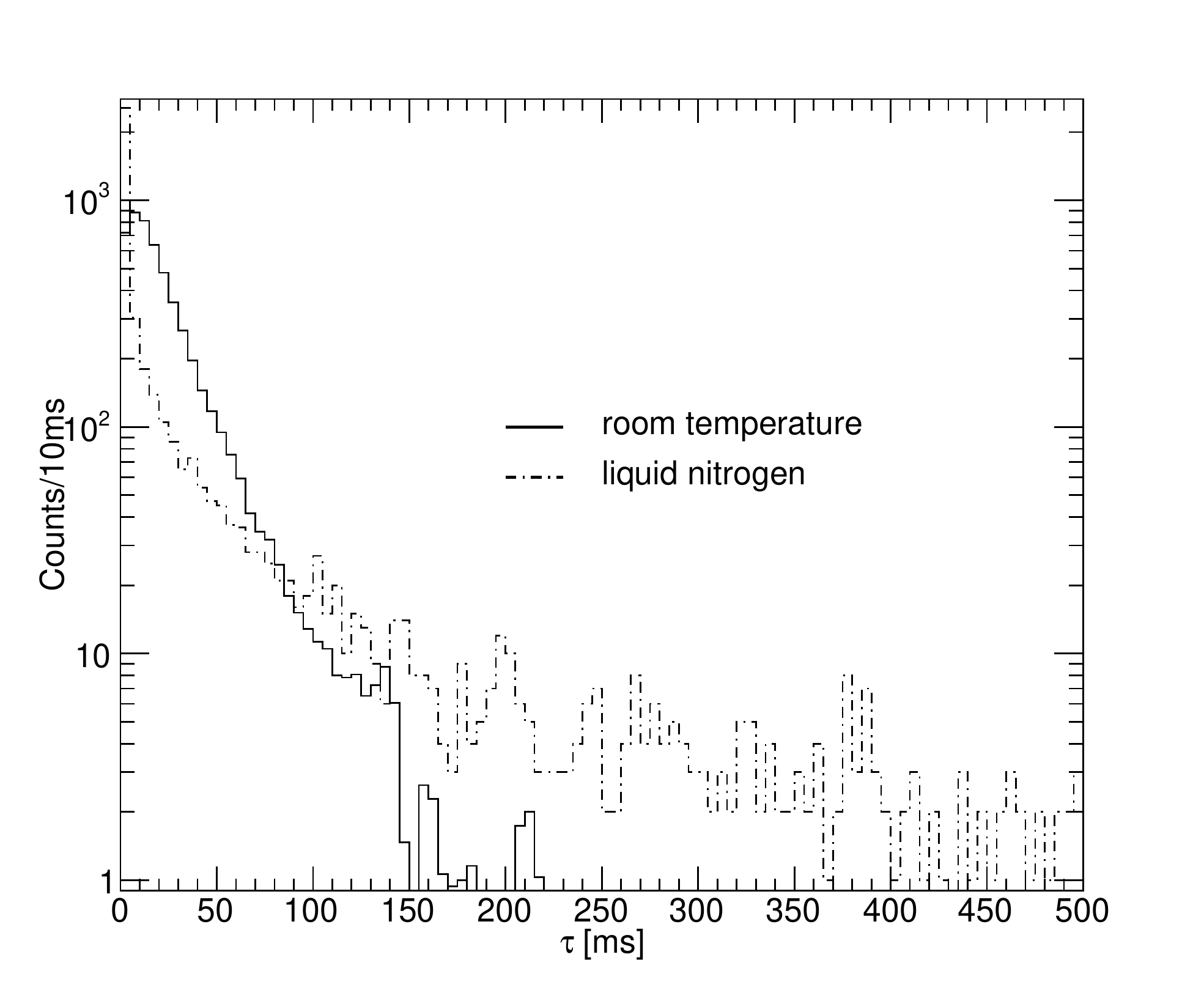}
			\centering
			\caption{Distribution of decay time constants of \textit{Topmetal\textrm{-}$II^{-}$}. At room temperature  $V_{reset}$ is 800 mV, and in liquid nitrogen $V^{'}_{reset}$ is 1.36 V. $V_{ref}$ is set to a fixed value of 618 mV. The number of pixels with a decay time constant of less than 3.3 ms (typical scan time) is about 500 in liquid nitrogen and 15 at room temperature, in the scan mode, We can not obtain the effective charge from these pixels.}	
			\label{decay constant of Topmetal-II-}		
		\end{figure}
			
		Since total pixel array in a sensor is connected to the same $V_{reset}$, uniformity of pixels at the same sensor is very important. The decay time constant distributions of total pixel array both at room temperature and in liquid nitrogen are shown in Figure. \ref{decay constant of Topmetal-II-}. The distribution tends to be broader in liquid nitrogen than at room temperature, while both cases show good uniformity among pixels. For our measurements, the pixels with a decay time constant less than 3.3 ms are dead pixels, the number of which is about 500. The presence of these dead pixels will have a bad effect on the accuracy of total charge. Therefore, in the design of next  \textit{Topmetal} sensors, we will improve the uniformity of decay time constants among pixels.
	
	\subsection{Noise Test}
\begin{figure} [htbp]
			\centering
			\includegraphics[width=1.0\columnwidth]{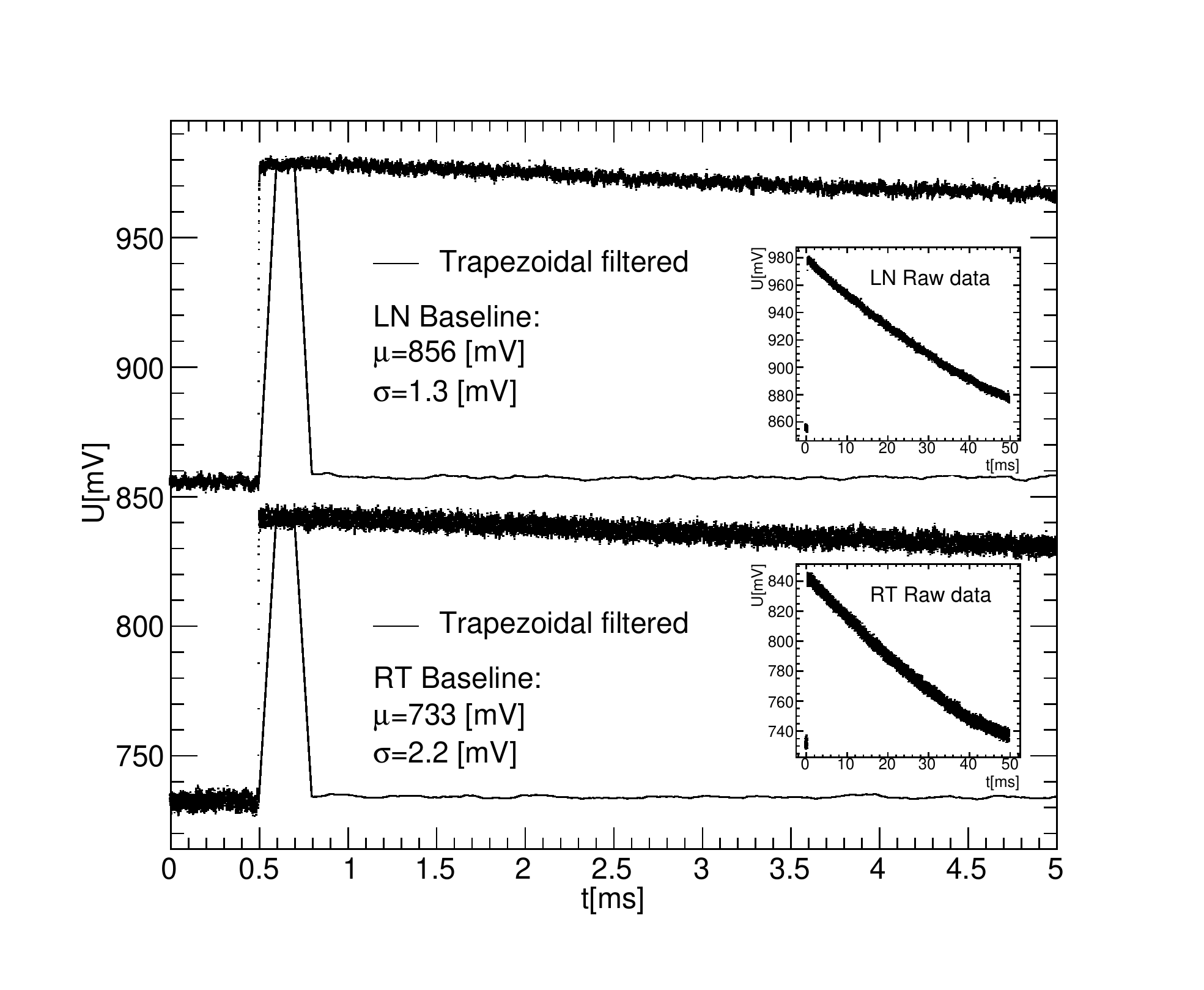}
			\centering
			\caption{CSA noise measurement, raw signal pattern and waveform after trapezoidal filtering. The decay time constants of the pixel at room temperature (RT) and in liquid nitrogen (LN) are adjusted to nearly the same. At room temperature, $V_{reset}$=800 mV, $V{ref}$=618 mV. When \textit{Topmetal-${II}^-$} is in liquid nitrogen, $V^{'}_{reset}$=1.36V, $V^{'}{ref}$=618 mV. }	
			\label{data of single pixel signal and after trapezoidal filter}		
		\end{figure}

\begin{figure} [htbp]
			\centering
			\includegraphics[width=1.0\columnwidth]{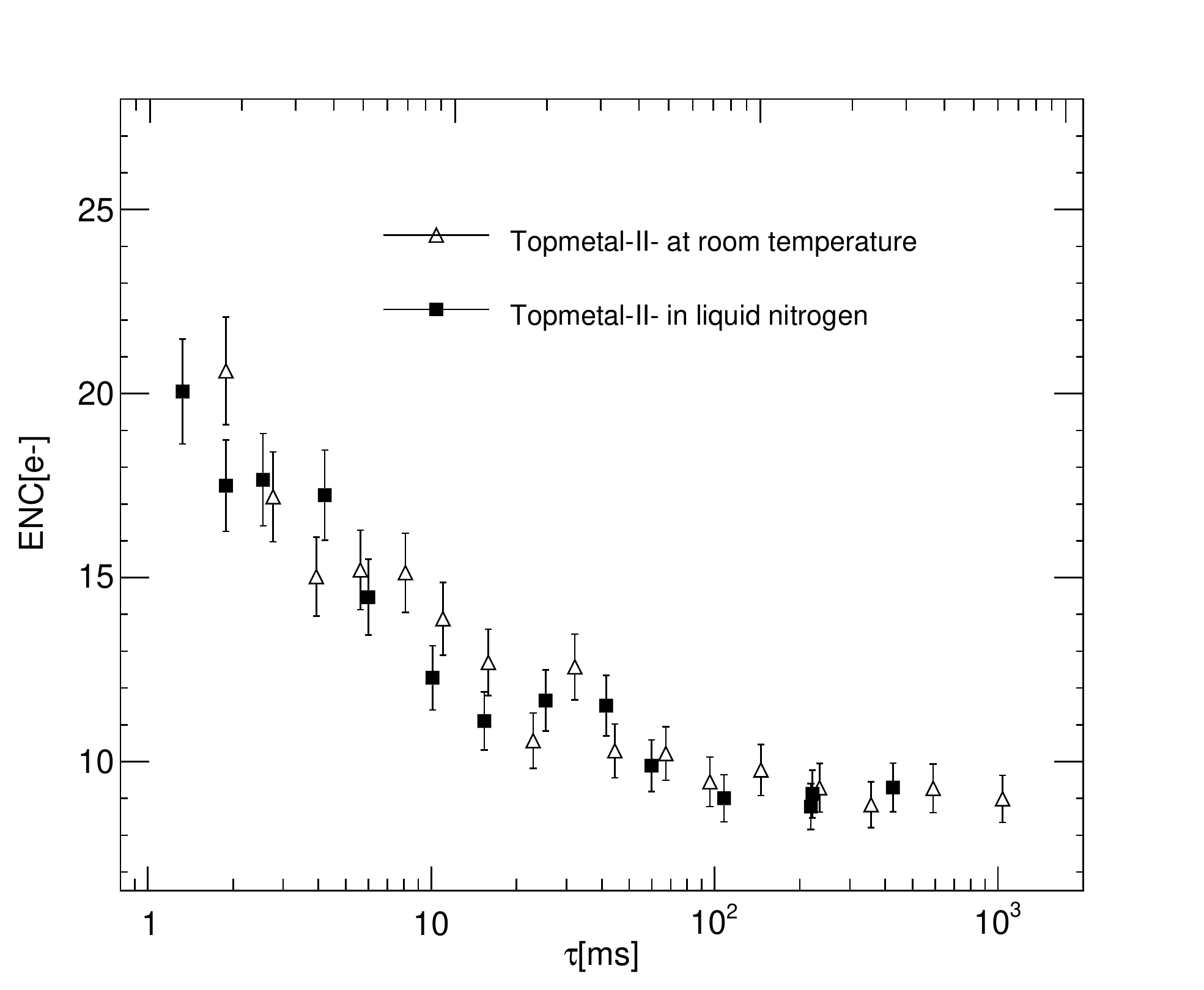}
			\centering
			\caption{A single pixel's decay time constant dependence of the noise of \textit{Topmetal-${II}^-$} at room tmeprature and in liquid nitrogen. We configure a reset voltage $V_{reset}$ for the sensor, and then measure the decay time constants and the amplitudes of a single pixel repeatedly to obtain the noise and the mean decay time constant. Each marker in this figure corresponds to a reset voltage  $V_{reset}$. The noise decreases with increased decay time constant. }	
			\label{ENC of single pixel of Topmetal-II}		
		\end{figure}

		A pulse of 200mV (peak-to-peak) is injected to guard ring as shown in Figure. \ref{data of single pixel signal and after trapezoidal filter}. The baseline voltage shifts about 123 mV from 733 mV to 856 mV, and the electronic noise is much lower ($\sigma$ 1.3 mV) in liquid nitrogen than that ($\sigma$ 2.2 mV) at room temperature. Since there is no shaper within \textit{Topmetal\textrm{-}$II^{-}$}, a digital trapezoidal filter\cite{digital_techniques_filter} is applied to the raw data to get the amplitude of signal and then the mean (${\mu})$ and standard deviation ($\sigma$) of amplitudes are measured. The equivalent noise charge (ENC) is calculated using the formula $ENC=C_d\times V_{injection} \times\sigma \slash \mu $.  We configure a reset voltage for a single pixel, and then measure repeatedly the decay time constants and the amplitudes to obtain the ENC and the  mean decay time constant. Figure. \ref{ENC of single pixel of Topmetal-II} shows the dependence of a single pixel ENC with the decay time constant from several milliseconds to one second. On the whole, the noise of a single pixel in liquid nitrogen is slightly smaller than those at room temperature.
	  
\begin{figure} [htbp]
			\centering
			\includegraphics[width=1.0\columnwidth]{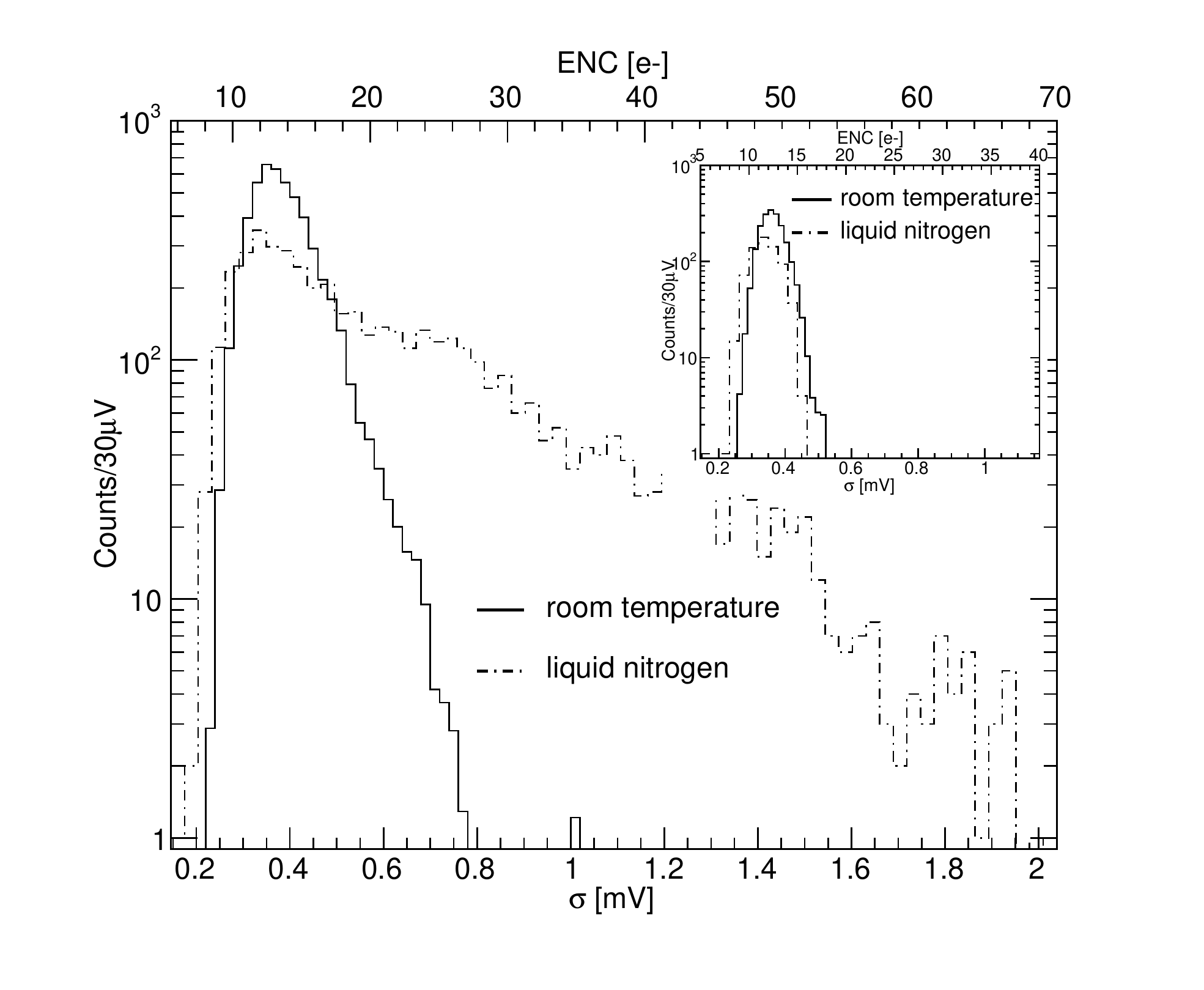}
			\centering
			\caption{ENC distrbution of Topmtal-II-. In liquid nitrogen, $V^{'}_{reset}$ is 1.36 V and the most probable values (MPV) of ENC among pixels is about 12 ${e^-}$. At room temperature, $V_{reset}$ is 800 mV and MPV is about 13 ${e^-}$. The small figure in the top right corner shows the noise distributions of the sensor in liquid nitrogen and at room temperature where the decay time constant is in the range of 10 to 50 ms, the mean (rms $\sigma$) value is 11.8 (1.4) and 12.6 (1.2) e$^-$ in liquid nitrogen and at room temperature respectively. }	
			\label{1d distribution of enc of Topmetal-II-}		
		\end{figure}

		The uniformity of ENC of \textit{Topmetal-${II}^-$} is significant, so we measured ENC of total pixel array using the same method described previously, setting $V_{reset}$ as 800 mV at room temperature and 1.36 V in liquid nitrogen. The ENC distribution in Figure. \ref{1d distribution of enc of Topmetal-II-} of total pixel array in liquid nitrogen is wider than that at room temperature. This is consistent with the measurement that decay time constants change in much larger range in liquid nitrogen for different pixels of \textit{Topmetal-${II}^-$}, since ENC is affected by the decay time constant when trapezoidal filter is applied to shape the signal. Most probable values (MPV) of ENC are 12 and 13 e$^-$ in liquid nitrogen and at room temperature respectively, while the mean ($\sigma$) value is 19.4 (10.4) and 13.5 (2.5) e$^-$. The small figure in the top right corner of Figure. \ref{1d distribution of enc of Topmetal-II-} shows that the ENC in liquid nitrogen is slightly less than that at room temperature when the decay time constant is in the range of 10 to 50 ms.
		
	\subsection{Linearity}
\begin{figure} [htbp]
			\centering
			\includegraphics[width=1.0\columnwidth]{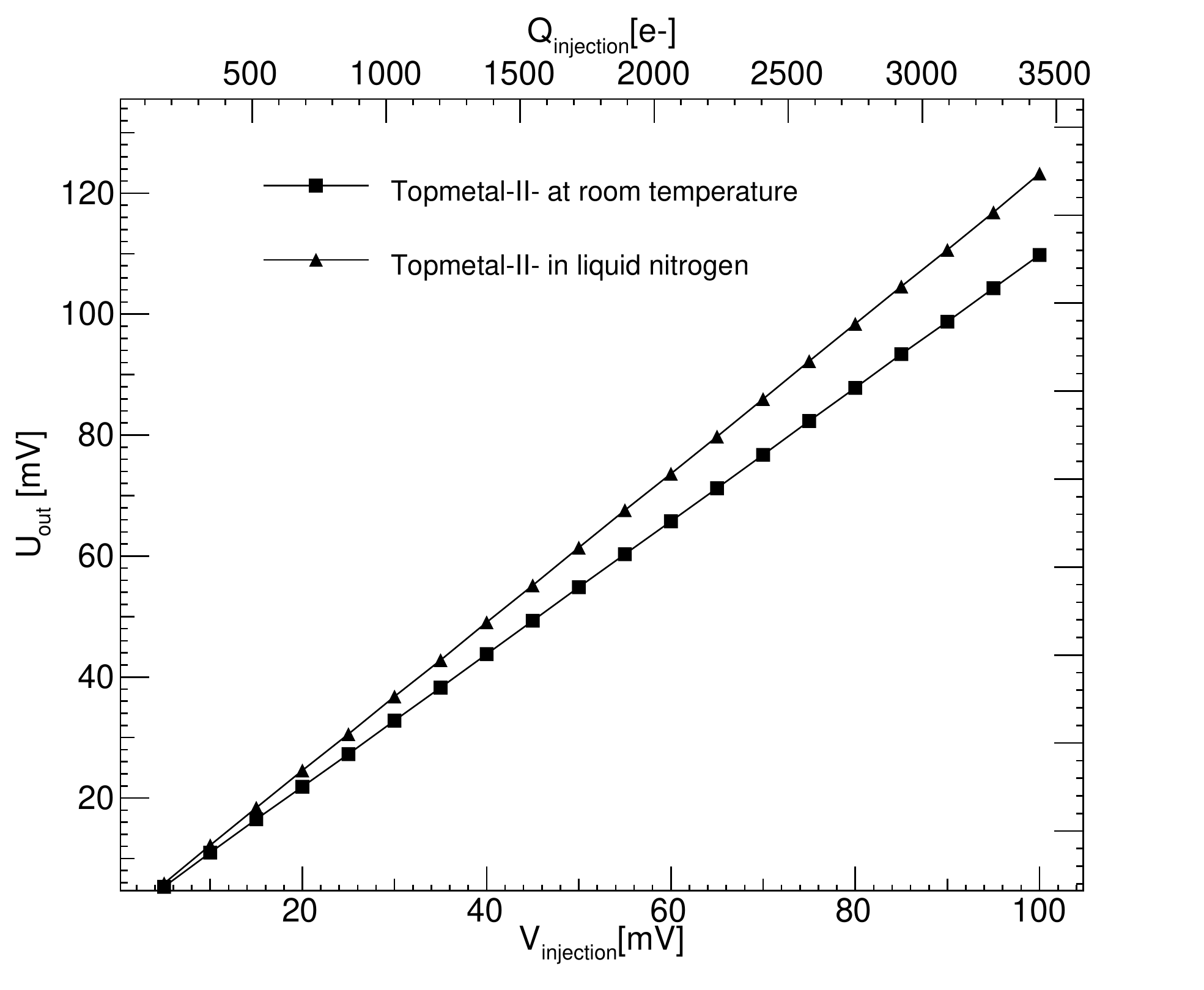}
			\centering
			\caption{Linearity of \textit{Topmetal-${II}^-$}. In liquid nitrogen, the amplitude is higher ($\sim$10\%) than that at room temperature for the same voltage of injected pulses. At room temperature, $V_{reset}$=800 mV, $V{ref}$=618 mV. When \textit{Topmetal-${II}^-$} is in liquid nitrogen, $V^{'}_{reset}$=1.36V, $V^{'}{ref}$=618 mV.}	
			\label{Linearity of Topmetal-II}		
		\end{figure}

		Linearity of \textit{Topmetal-${II}^-$} is of great importance to get injected charges. It has been measured and compared under both conditions of room temperature and liquid nitrogen. The voltage of pulses injected to guard ring varies from 10 mV to 100 mV with a step voltage of 10 mV. In Figure. \ref{Linearity of Topmetal-II}, we can see that the linearity of \textit{Topmetal-${II}^-$} is excellent. In liquid nitrogen, the amplitude is higher ($\sim$10\%) than that at room temperature for the same voltage of injected pulses.
				
\section{Summary}
	We demonstrate that \textit{Topmetal-${II}^-$} can operate well in liquid nitrogen. The most probable value of ENC of the sensor is about 13 e$^{-}$ at room temperature and 12 e$^{-}$ in liquid nitrogen after a digital trapezoidal shaper is applied. The ENC has a much wider spread in liquid nitrogen than that at room temperature because of the wider spread decay time constants. The sensor has a linear response to injected pulse signal. This \textit{Topmetal-${II}^-$} is a promising candidate to be applied in a liquid argon or xenon TPC for low rate experiments without charge amplification needed. We will explore applications based on \textit{Topmetal-${II}^-$} at cryogenic temperature.
	 
	 Also in the design of the next version of \textit{Topmetal} sensors for the purpose of a cryogenic temperature TPC, we will further improve the uniformity of decay time constants among pixels and reduce the noise of charge sensitive amplifier.
\section*{Acknowledgments}
	This work is supported by the National Natural Science Foundation of China under Grant No.11375073, No.11305072, and No.1232206.


\end{document}